\documentclass[a4paper, 11pt, twocolumn]{article}
\usepackage[utf8]{inputenc}
\usepackage[T1]{fontenc}
\usepackage{graphicx}
\usepackage{longtable}
\usepackage{wrapfig}
\usepackage{rotating}
\usepackage[normalem]{ulem}
\usepackage{amsmath}
\usepackage{amssymb}
\usepackage{capt-of}
\usepackage{hyperref}
\usepackage{cleveref}
\usepackage{cuted}
\usepackage{breqn}

\usepackage{microtype} 

\usepackage{amsmath,amsfonts,amsthm} 

\PassOptionsToPackage{svgnames}{xcolor}
\usepackage{awesomebox}
\usepackage{cleveref}

\usepackage[size=small, labelfont=bf, up, textfont=it]{caption} 

\usepackage{booktabs} 

\usepackage{lastpage} 

\usepackage{graphicx} 

\usepackage{enumitem} 
\setlist{noitemsep} 

\usepackage{sectsty} 
\allsectionsfont{\usefont{OT1}{phv}{b}{n}} 

\usepackage{geometry} 

\usepackage[T1]{fontenc} 
\usepackage[utf8]{inputenc} 

\usepackage{XCharter} 

\usepackage{fancyhdr} 
\fancypagestyle{firstpage}{ 
	\fancyhf{}
}

\usepackage{titling} 

\newcommand{\HorRule}{\color{DarkGoldenrod}\rule{\linewidth}{1pt}} 

\pretitle{
	\vspace{-30pt} 
	\HorRule\vspace{10pt} 
	\fontsize{32}{36}\usefont{OT1}{phv}{b}{n}\selectfont 
	\color{DarkRed} 
}

\posttitle{\par\vskip 15pt} 

\preauthor{} 

\postauthor{ 
	\vspace{10pt} 
	\par\HorRule 
	\vspace{20pt} 
}

\usepackage{amsmath}

\usepackage{lettrine} 
\usepackage{fix-cm}	

\newcommand{\initial}[1]{ 
	\lettrine[lines=3,findent=4pt,nindent=0pt]{
		\color{DarkGoldenrod}
		{#1}
	}{}%
}
\usepackage{xstring} 

\newcommand{\lettrineabstract}[1]{
	\StrLeft{#1}{1}[\firstletter] 
	\initial{\firstletter}\textbf{\StrGobbleLeft{#1}{1}} 
}

\newcommand{\PaperTitle}[1]{\def\@PaperTitle{#1}}
\newcommand{\Archive}[1]{\def\@Archive{#1}}
\newcommand{\Authors}[1]{\def\@Authors{#1}}
\newcommand{\JournalInfo}[1]{\def\@JournalInfo{#1}}
\newcommand{\Abstract}[1]{\def\@Abstract{#1}}
\newcommand{\Keywords}[1]{\def\@Keywords{#1}}

\usepackage{titling,kantlipsum}

\renewenvironment{abstract}
 {\small
  \vspace{-100pt}
  \begin{center}
  \bfseries \vspace{-.5em}\vspace{0pt}
  \end{center}
  \list{}{%
    \setlength{\leftmargin}{5mm}
    \setlength{\rightmargin}{\leftmargin}%
	\setlength{\topmargin}{0mm}
  }%
  \item\relax}
 {\endlist}

\usepackage[backend=bibtex,style=authoryear,natbib=true, style = nature]{biblatex} 

\usepackage[autostyle=true]{csquotes} 

\DeclareSourcemap{
  \maps{
    \map{
      \step[fieldsource=url,
            match=\regexp{/books\.google\.},
            fieldset=url, null]
    }
  }
}

\usepackage[english]{babel}
\author{Casper van Elteren\(^{0, 1, 2, 3}\), Vítor V. Vasconcelos\(^{1, 3}\), and Mike Lees\(^{1, 2, 3}\)}
\date{}
\title{Criminal organizations exhibit hysteresis, resilience, and robustness by balancing security and efficiency}
\begin{document}

\twocolumn[
 \begin{@twocolumnfalse}
 \maketitle
 \begin{abstract}
\lettrineabstract{The interplay between (criminal) organizations and (law enforcement) disruption strategies is critical in criminology and social network analysis. Like legitimate businesses, criminal enterprises thrive by fulfilling specific demands and navigating their unique challenges, including balancing operational visibility and security. This study aims at comprehending criminal networks' internal dynamics, resilience to law enforcement interventions, and robustness to changes in external conditions. Using a model based on evolutionary game theory, we analyze these networks as collaborative assemblies of roles, considering expected costs, potential benefits, and the certainty of expected outcomes. Here, we show that criminal organizations exhibit strong hysteresis effects, with increased resilience and robustness once established, challenging the effectiveness of traditional law enforcement strategies focused on deterrence through increased punishment. The hysteresis effect defines optimal thresholds for the formation or dissolution of criminal organisation. Our findings indicate that interventions of similar magnitude can lead to vastly different outcomes depending on the existing state of criminality. This result suggests that the relationship between stricter punishment and its deterrent effect on organized crime is complex and sometimes non-linear. Furthermore, we demonstrate that network structure, specifically interconnectedness (link density) and assortativity of specialized skills, significantly influences the formation and stability of criminal organizations, underscoring the importance of considering social connections and the accessibility of roles in combating organized crime. These insights contribute to a deeper understanding of the systemic nature of criminal behavior from an evolutionary perspective and highlight the need for adaptive, strategic approaches in policy-making and law enforcement to disrupt criminal networks effectively.} 
\end{abstract} \end{@twocolumnfalse}]

\footnotetext[0]{Corresponding author: c.vanelteren@uva.nl}
\footnotetext[1]{Computational Science Lab, Informatics Institute, University of Amsterdam, The Netherlands}
\footnotetext[2]{Institute for Advanced Study, University of Amsterdam, Amsterdam, The Netherlands}
\footnotetext[3]{POLDER Center, Institute for Advanced Study, The Netherlands}



Crime pays. The United Nations Office on Drugs and Crime estimates that transnational criminal organizations generate an astonishing annual revenue of approximately \$1.6 trillion \cite{zotero-24804}, equivalent to Canada's Gross Domestic Product. This staggering financial success places criminal organizations among the wealthiest and most influential entities globally, underscoring the substantial challenges law enforcement agencies and governmental bodies face in combating crime.

Beyond its financial prowess and undue influence, crime exerts other profound social, health, and environmental impacts. The illicit drug trade contributed to escalated drug use and disorders, reflecting broader societal issues. The decade ending in 2021 saw a 23\% increase to over 296 million people using drugs and a 45\% increase in drug use disorders to 39.5 million individuals \cite{UNODC2023}. These trends reflect the direct consequences of the drug trade but also highlight the resulting social and economic inequalities, with significant healthcare disparities in treatment access. The impact is acute among the youth, especially in Africa \cite{UNODC2023}. 
Furthermore, the drug economy fuels additional criminal activities such as illegal logging and mining, particularly in the Amazon Basin, leading to environmental degradation and human rights violations. Indigenous communities and other minority groups are disproportionately affected by these illegal activities. Finally, drug trade funds non-state armed and insurgency groups, contributing to regional instability and potentially escalation to global crises \cite{UNODC2023}.

The success and far-reaching damaging impact of criminal organizations highlight the need to understand the conditions under which criminal organizations thrive or falter \cite{Finckenauer2005,Hagan2006}. Criminal organizations exhibit intriguing similarities to legitimate businesses. Like successful licit enterprises that cater to societal needs, criminal organizations function by fulfilling a specific demand \cite{Schelling2023}. Criminal organizations bring illegal products to market through international supply chains, such as cocaine, through a network of coca farmers, smugglers, runners, and dealers \cite{Davila2021}. Running an illicit operation requires goal-orientated and management skills not uncommonly found in successful businesses \cite{Bunt2014, Schelling2023}.

Legitimate organizations thrive on market visibility to attract customers and collaborators, whereas criminal organizations must balance visibility with the risk of exposure to law enforcement. This challenge, known as the security-efficiency trade-off in criminal literature, requires criminal entities to navigate between operational effectiveness and security  \cite{Morselli2007,Morselli2009a}. To achieve this, criminals rely on “dark” social networks to facilitate covert cooperation, essential for their operation \cite{vonLampe2004, Nevala2004}. The structure of these networks is pivotal; networks with denser connections and a smaller average maximum distance between any two nodes are more efficient in communication, aiding in the growth and functionality of the organization. Conversely, secure networks maintain higher social distance between members, reducing the risk of complete exposure when a member is compromised \cite{Calderoni2020, Bouchard2020a, Ayling2009, Diviak2023, Duijn2016}. The security-efficiency trade-off is, thus, a spectrum that characterizes networks based on their connectivity structure. The operational conditions and types of social networks employed by these organizations significantly influence their likelihood of success and ability to sustain operations under various conditions \cite{Duijn2016, Duijn2014a}.

The increasing emphasis on structural analysis based on criminal reports has led to the idea of finding a structural fingerprint to improve interventions by law enforcement \cite{Duijn2016, Duijn2014a}. Social networks of criminals, commonly constructed from biased data sources such as arrest records, telephone records, or informant data, illuminate divergent facets, engendering distinct network structures \cite{Nevin2023}. This inherent bias can distort evaluation of when a network is considered secure or efficient. Criminal networks often exhibit patterns where not all nodes are connected, resulting in structures characterized by dense connections within communities, and sparser connections across different communities \cite{Bouchard2020a, Reuter2020, Ficara2022}. For example, the 'Ndrangheta network \cite{legramanti2022extended} displays a nested and hierarchical structure, with bosses and affiliates forming coherent sub-groups within local units. This structure suggests a sophisticated organization and division of roles within criminal networks, which are crucial for understanding the operational dynamics of criminal networks.

While complex network analysis has gained traction in studying crime data, its application in predicting static and dynamic properties of criminal networks is still a relatively unexplored area \cite{lopes2022machine, Ahmadi2023, Assouli2021}. Machine learning methods can effectively anticipate future criminal associations given the current system \cite{lopes2022machine} but do not provide a clear picture of the mechanisms leading to the formation and collapse of criminal organizations in different environments, a gap our research seeks to address.
Regrettably, scant attention has been devoted to exploring the intricate interplay of individuals' dynamic decisions and their social network.

At the core of criminal organizations are individuals who evaluate the anticipated cost and benefit of a criminal act, guided by a certainty of the expected payout \cite{Schelling2023, Finckenauer2005, Bunt2014, Duijn2016, Calderoni2022}. Successful execution of a criminal activity often hinges on collaboration among a specific set of distinct roles.

This study explores the underlying mechanisms that govern criminal network formation, expansion, and potential dissolution. Using a model based on concepts from evolutionary game theory, we analyze these networks as collaborative assemblies of roles, considering expected costs, potential benefits, and the certainty of expected outcomes. Our experiments investigate how the interplay of dynamics on social networks and changes to environmental conditions impact the robustness of criminal organizations against law enforcement efforts. The approach complements the extensive literature on multi-actor coordination games by applying them to a criminal context \cite{Cooper1999, Weidenholzer2010, Cooper2020, Encarnacao2016, Raducha2022, Raducha2023, Broere2017, Weinans2024, Berenji2014, Cressman1998}. 

Our results demonstrate an important effect of hysteresis (path dependency) in which the best intervention (e.g., increased punishment) and its effect depends on whether a community already has high levels of criminality or not. We explore the ability of the criminal organization to withstand shocks (sudden loss criminals) --- which we call resilience --- and its ability to withstand changes in external conditions --- which we call robustness. We identify a critical cost-to-benefit ratio where the formation or dissolution of a criminal organization is the most susceptible to both external perturbations, e.g. law enforcement disruption, and internal perturbations, e.g. availability of criminals. Importantly, network effects play a pivotal role in enhancing the robustness and resilience of criminal organizations. Criminal organizations reach peak robustness and resilience, when access to specialized roles necessary to form the organization maximizes, thereby cultivating an environment conducive to exploiting criminal opportunities.

\section{Modeling criminal organizations}
\label{sec:org53595a9}
An organization, whether licit or illicit, succeeds when all required roles coordinate in the same strategy: criminal or non-criminal. If an agent decides to engage in a criminal act, they receive a payout only if the other members of the organization also choose to perform a criminal act, which results in receiving a benefit; they also incur a cost, embodying the chance of getting detected by law enforcement, or other environmental costs such as loss of shipment or influence of rivaling gangs (see for more background \ref{sec:org2890a8a}).

The agents attempt to form a complete organization, a full set of roles, using their social network.
Across different networks, reflecting the spectrum of efficiency-secrecy trade-off, we study the emergence of criminal-dominated states and their stability to shocks in criminal numbers --- resilience --- and to changes in external conditions --- robustness.

We consider a population of size \(Z\). Each agent, $a_i = \{ r_i, s_i\}, i \in \{1,...,Z\}$, is characterized by having one of $R$ fixed roles, $r_i\in\{1,...,R\}$, and dynamic state, $s_i\in\{0,1\}$ . The fraction of each role is denoted as \(z_r=Z_r/Z\), satisfying the constraint \(\sum_{r=1}^R z_r = 1\). For instance, with \(R = 3\), a (criminal) organization consists of three roles: e.g., distribution, production, and management. A state \(s_i=0\) represents a non-criminal state and \(s_i=1\) a criminal state of agent \(i\).

\subsection{Strategy update}
\label{sec:org52a18cc}

Agents consider the alternative states by exploring their neighborhood and considering forming an organization with neighbors of complementary roles \(n\) times. An agent $a_i$ is connected with an agent $a_j$ when $g_{ij}=1$; otherwise, $g_{ij}=0$. $G=[g_{ij}]$ forms an undirected, unweighted adjacency matrix, representing the graph of potential interaction partners.

Forming an organization entails finding other agents with all the complementary roles and coordinating to engage in criminal or non-criminal activity. Formally, agent \(a_i\) samples $R-1$ agents, one from each set $\{a_k: r_k=r\;\text{and}\; a_{ik}=1\}$, where $r\in\{1,...,R\}\setminus r_i$, creating a random set $M_i$ of $R-1$ complementary-role agents in the neighborhood of $a_i$ to interact with. 

In each game, the agent interacts with the set $M_i$ and receives a benefit \(b\) when all $R$ individuals are criminals. Criminal activity, however, entails a cost \(c\). The payout for each of these interactions of agent $i$ in state $s\in\{0,1\}$ is given as

\begin{equation}
\label{payout}
\pi_s^{(i)} = s \left(b \prod_{s_j \in M_i} s_j - c\right).
\end{equation}

The agent aims to maximize their payoff by considering the expected payoff of each strategy by averaging the payoff of the interactions with \(n\) random $M_i$ groups, $\langle\pi^{(i)}_{s}\rangle$.

In each time step, a random agent, $i$, is selected to consider changing their strategy. The probability of changing one's state from criminal, $s=1$, to non-criminal, $s=0$, and vice versa is computed based on the expected payout difference of each strategy. The probability of switching from strategy \(s\) to a new strategy \(s'\) is given by

\begin{equation}
\label{fermi}
p_{s \to s'} = \frac{1}{1 + e^{-\frac{1}{\epsilon}\left(\langle\pi_{s'}^{(i)}\rangle - \langle\pi_{s}^{(i)}\rangle\right)}},
\end{equation}
where \(\epsilon\) represents the decision error controlling the agent's uncertainty when changing strategy. Notice that the transition or conditional probability of changing to a given strategy is independent of the base state, mapping directly to a multinomial logit model that reflects a law of relative effect \cite{Helbing2010} and, for a single role, maps to the classic Glauber dynamics for the kinetic Ising models \cite{Glauber1963}.

In the extreme case where \(\epsilon \to \infty\), using \cref{fermi}, an agent will choose their next state independently of the information it receives from their neighbors with uniform probability. Conversely, as \(\epsilon \to 0\), any minimum expected payoff difference will lead to a change of strategy.

\section{Results}
\subsection{Dynamics of efficient criminal organizations}
\label{sec:orga2bf2e6}
Efficient communication in a system implies that each agent in the system can, in principle, communicate with all other agents in
the system, i.e., $g_{ij}=1$ for all $i$ and $j$. For large populations, the dynamics of the collective can be determined by considering the fraction of criminals in a particular role, denoted \(x_r=\frac{1}{z_r Z}\sum_{i: r_i=r}s_i\). In the limit of \(n\to \infty\), i.e., perfect estimation of the average payoff with the current neighborhood, the system dynamics will follow (see \ref{sec:org28cfbc4} for details)

\begin{equation}
\label{limiting_change}
\begin{split}
 \lim_{n\to\infty} \frac{d}{dt} x_r = \frac{1}{1 + e^{- \frac{1}{\epsilon} b(\prod_{q = 0, q \neq r}^R x_q - \frac{c}{b})}} - x_r,
\end{split}
\end{equation}

which distinctly shows how the emergent instantaneous change for the collective of individuals with a given role is controlled  (\cref{stability_plot}).

The parameters \(b\), \(c\), and \(\epsilon\) control different aspects of how socially viable the criminal strategy is for the efficiently communicating community.
The equation highlights three dependencies: the cost-to-benefit ratio, $c/b$, the error-to-benefit, $\epsilon/b$, and the number of roles, $R$.

For the upcoming simulations, we will set \(R=3\), where each role's occurrence is equal, i.e., \(z_1 = z_2 = z_3\), unless otherwise specified. Additionally, the cost- and error-to-benefit ratios, $c/b$ and $\epsilon/b$, will be manipulated by maintaining \(b = 1\) and adjusting \(c\) and $\epsilon$. This choice stems from the dual impact of \(b\) on both the curvature of the decision curve as well and the
cost-to-benefit ratio (see \cref{limiting_change}). The effect of $R$ on the dynamics are qualitatively similar for $R \ge 2$ and additional results for different $R$ can be found in \cref{sec:orgc0093f7}.

\cref{stability_plot} shows the fixed points of the dynamics of efficient criminal organizations, which characterize where the system is attracted to (stable fixed points) or repelled from (unstable fixed points). In the case of high decision error, high $\epsilon$, the system exhibits a single stable fixed point, which attracts the dynamics to an intermediate fraction of criminals. In this regime with a single equilibrium, increasing the cost-to-benefit ratio consistently and reversibly decreases the fraction of criminals. When the decision increases further, all dynamics resort to a random choice regardless of the cost of the criminal act, e.g., compare \(\epsilon = \infty\) to \(\epsilon = 5\). This reflects the conditions in which the lines between criminal and non-criminal acts are blurred. An example could be the formation of a "second" government such as the prevalence of the Mafia in southern Italy \cite{schneider2003reversible}. However, reducing the decision error can either decrease the fraction of criminals if the cost-to-benefit ratio is high enough or increase it for a low cost-to-benefit ratio. The number of roles, $R$, reflecting the complexity of the task, controls the critical cost-to-benefit ratio that determines the direction of the effect (precisely, $(c/b)^{\mathrm{crit}}=1/2^{R-1}$). 
For low decision error, low $\epsilon$, the dynamics may exhibit three fixed points: two stable attractors, one with a high fraction of criminals (e.g., the points between A and B) and one with low (e.g., the points between C and D), and an unstable state at an intermediate fraction of criminals. For a high cost-to-benefit ratio, above $(c/b)^{\mathrm{H}}$, only the attractor with a low fraction of criminals is present; for low cost-to-benefit ratio, below $(c/b)^{\mathrm{L}}$, only the attractor with a high fraction of criminals is present; and for intermediate values, between $(c/b)^{\mathrm{L}}$ and $(c/b)^{\mathrm{R}}$, the three are present and the dynamics can be attracted to two different stable states. When the system is in a stable state with a high fraction of criminals (A), changes to the cost-to-benefit ratio are largely ineffective in changing the fraction of criminals ($A \to B$) unless they surpass $(c/b)^{\mathrm{R}}$ ($B \to C$). Similarly, when the system is in a stable state with a low fraction of criminals (C), a reduction of the cost-to-benefit will not lead to an increase ($C \to D$) unless the reduction is beyond $(c/b)^{\mathrm{L}}$ ($D \to C$). 
In the regime where the three states are present, with intermediate values of cost-to-benefit ratio, the initial number of criminals determines where the system is attracted. The unstable fixed point determines the critical mass of criminals below which criminal activity collapses to the low criminal state and above which it grows to the high criminal state. As the cost-to-benefit ratio increases, this critical mass also increases.


\subsection{State-dependent fortitude: hysteresis in criminal organizations}
\label{sec:org783a100}
The ability to rebound from interventions is crucial for the persistence of criminal organizations. These interventions can be categorized into two dimensions.

\begin{figure}[htbp]
\centering
\includegraphics[width=.9\linewidth]{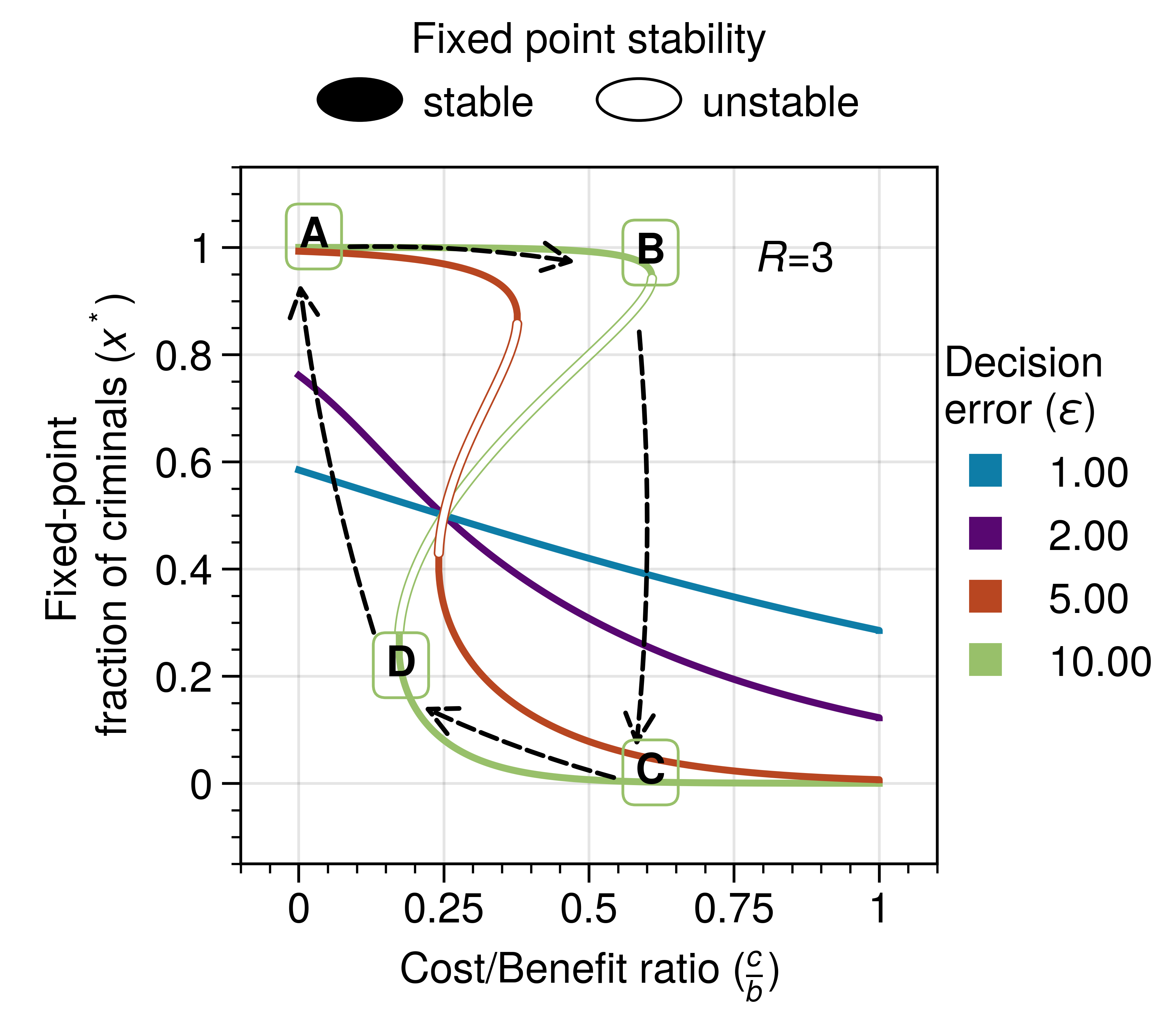}
\caption{\label{stability_plot}Modulating the decision error induces hysteresis in the system's behavior. In scenarios of low cost-to-benefit and elevated decision error, a single stable attractor prevails. However, as the decision error diminishes, an unstable attractor surfaces, acting as a threshold representing the minimum required fraction of criminals needed to saturate the market. The criminal strategy materializes only under conditions of low cost-to-benefit and recedes with increasing costs. Importantly, in responds to an intervention, the system shows increased robustness and resilience when formed under these conditions. For a comprehensive exploration of bifurcation patterns based on the number of roles, refer to \ref{sec:orgc0093f7}}
\end{figure}

Firstly, we consider fluctuations in the fraction of criminals, focusing again on \cref{stability_plot}. Over time, the number of criminals may change due to various factors, such as deaths, rivalries, and other dynamics. The ability to resist fluctuations in the population will be denoted as \emph{resilience} \cite{Holling2013, Ayling2009}.

\begin{figure}[th!]
\centering
\includegraphics[width=.9\linewidth]{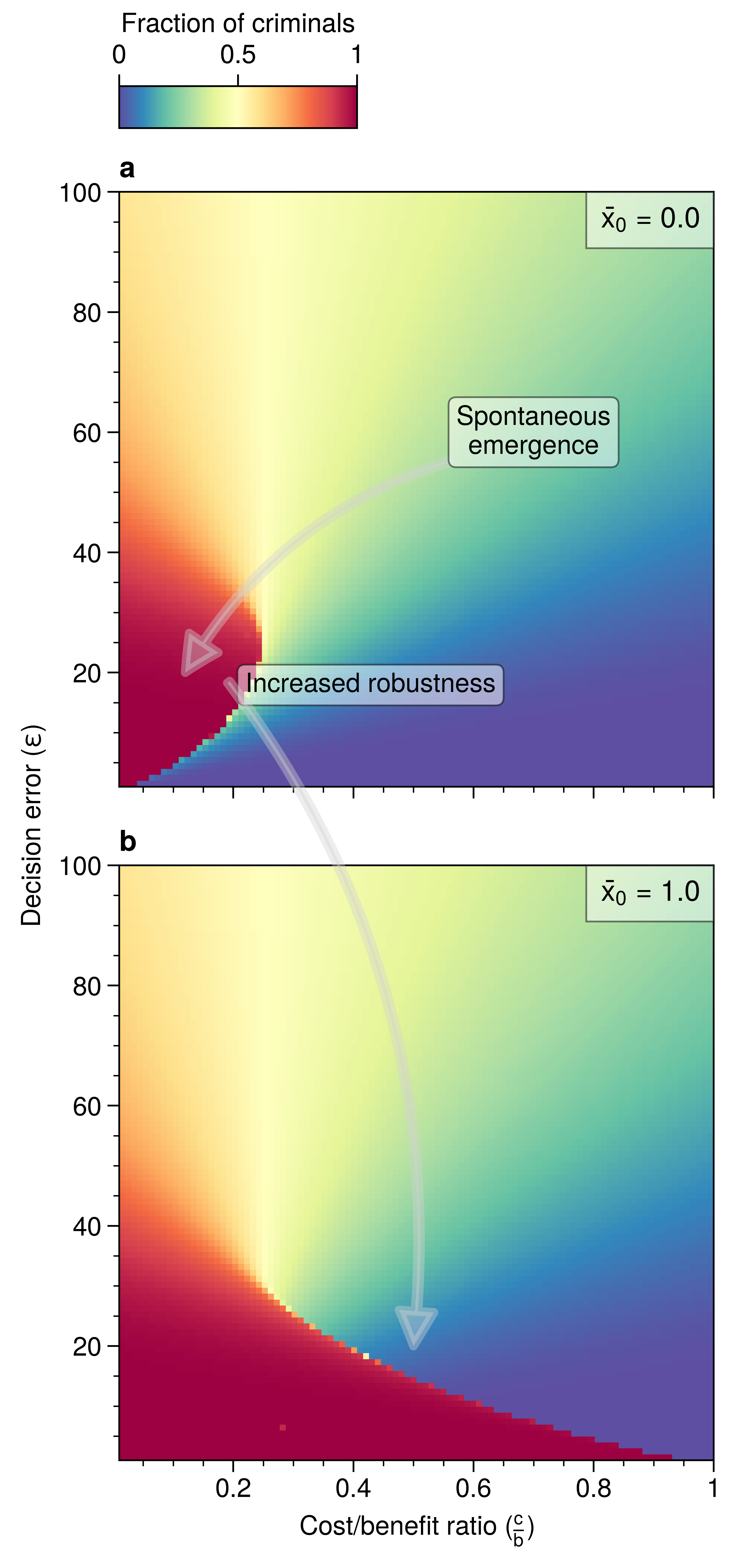}
\caption{\label{phase_plot}The emergence of criminal organizations depends on the decision error, the cost-to-benefit ratio, and the initial fraction of criminals. (a) For low levels of crime in society, criminal organizations spontaneously emerge when the benefit outweighs the cost, and the decision error (\(\epsilon\)) is moderately high. (b) The stability region for criminal organizations becomes large for higher initial fraction of crime, indicating that criminal organizations are more robust to decision error. Once a criminal organization is formed, its \emph{resilience} increases; consider the initial conditions with a lack of criminals in society (a), once the criminal organization forms, the criminal \emph{resilience} is higher. That is, stronger interventions are needed to disrupt the criminal organization.}
\end{figure}

Secondly, the viability of a criminal market is influenced by the relative market value of the product being sold and different risks. Fluctuations in supply and demand and law enforcement surveillance impact the sustainability of criminal organizations. The ability to resist fluctuations due to external pressure is denoted as \emph{robustness}\cite{Holling2013,Ayling2009}. For this, we will use \cref{phase_plot}.

As described above, at high decision error, there is a single attractor and, thus, the system is fully resilient and robust. Perturbations to the number of criminals, including recruitment actions, will lead to temporary changes in the number of criminals but the same end result. This is seen by the single stable point in \cref{stability_plot} and, in \cref{phase_plot}, by realizing the identical sections of the two panels, where both extreme initial conditions lead to the same final fraction of criminals across that whole range of parameters (in the figure this includes the whole $\epsilon \gtrsim 40$ region). 

A critical regime can be identified at low decision error with the emergence of the two stable states described above with high and low numbers of criminals. The state with many criminals (top solid lines in \cref{stability_plot} and bottom panel in \cref{phase_plot}) is resilient to interventions in criminal activity that do not bring the numbers below the unstable state in \cref{stability_plot}. Besides, that criminal state is robust to changes in the cost-to-benefit ratio and decision error (the red region in \cref{phase_plot}). At the border of the red region, however, the highly criminal state loses its resilience and robustness. 

For instance, consider the dynamics of a criminal organization with a decision error \(\epsilon = 5\) in \cref{stability_plot}, corresponding to the red area in \cref{phase_plot}b. When the criminal organization is already established (\(A\)), the cost-to-benefit ratio must increase until \(\frac{c}{b} \simeq 0.56\), at which point the criminal organization becomes highly sensitive to perturbations in both the number of criminals and the cost-to-benefit ratio. The criminal organization cannot recover from the perturbations then, heading to the non-criminal state (\(B \to C\) transition).
%
Once the criminal organization is dissolved, $B$ (also the corresponding blue area in \cref{phase_plot}a), reducing the cost-to-benefit ratio or slightly increasing the fraction of criminals does not reverse the effect. Instead, the system remains non-criminal even as costs to criminal activities are reduced or benefits increase (\(C \to D\)). At low enough costs, the criminal organization emergences (\(D \to A\) transition and red area in \cref{phase_plot}a).

\begin{figure}[htbp]
\centering
\includegraphics[width=.9\linewidth]{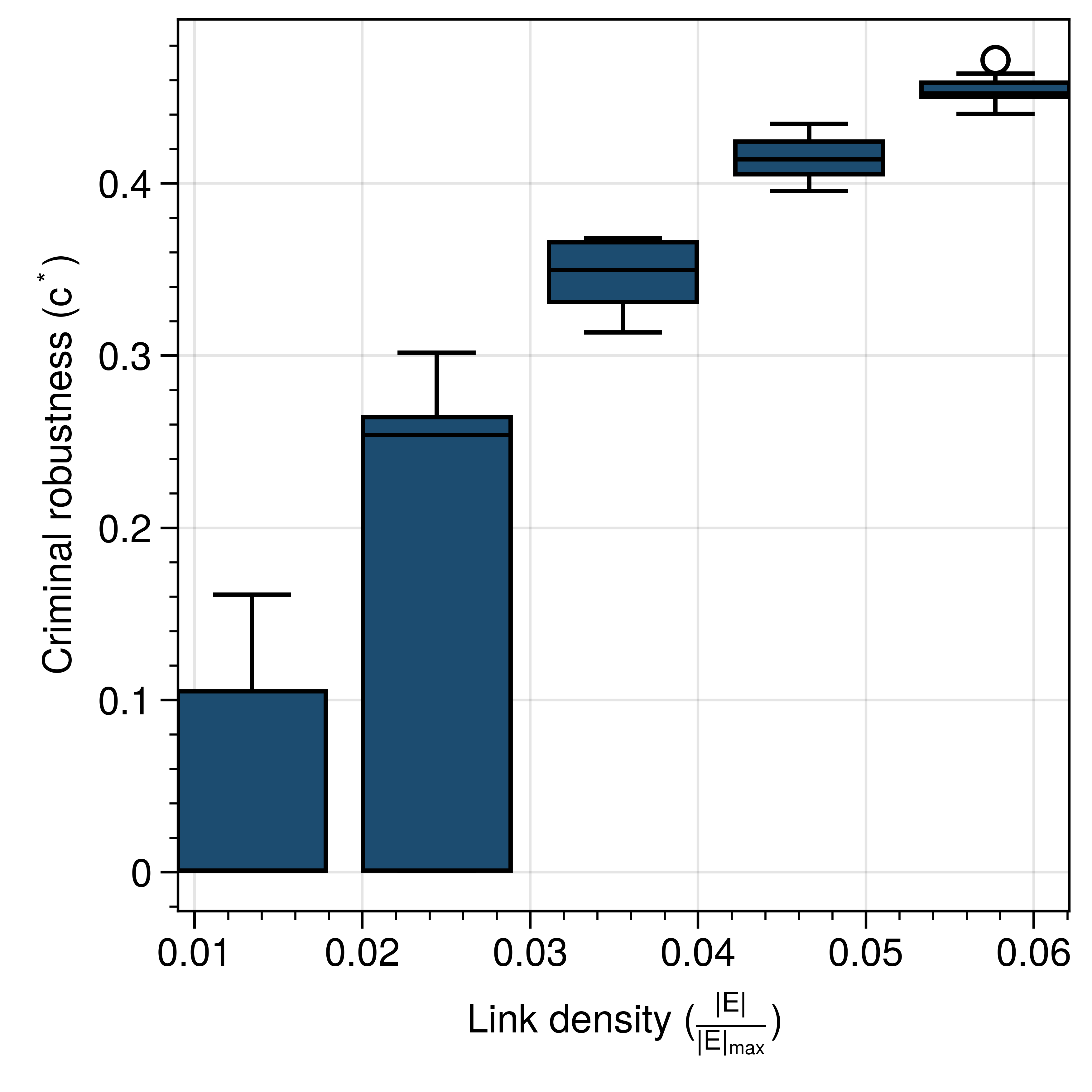}
\caption{\label{link_plot}Link density promotes the \emph{resilience} of criminal organizations. The interquantile range is visualized for graphs with role assortativity <= -0.5 and \(\epsilon = 10\). Outliers are represented by un-filled scatters when they exceed 1.5 the interquantile range.}
\end{figure}

\begin{figure*}[ht]
\centering
\includegraphics[width=.9\linewidth]{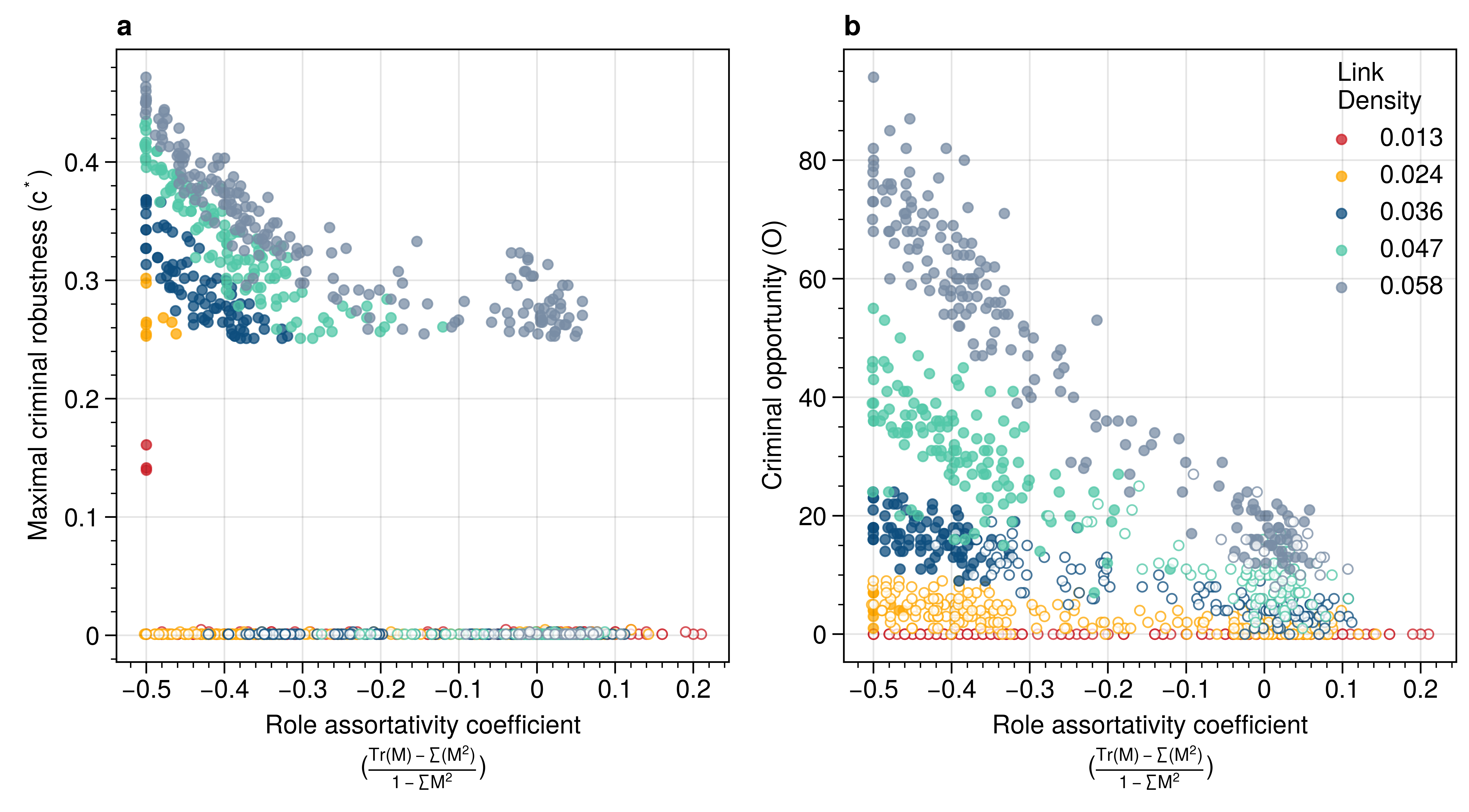}
\caption{\label{assort_plot}The robustness of criminal organizations is promoted by networks characterized by elevated density and dissortativity. In panel (a), an inverse relationship is observed, wherein increasing assortativity leads to a decline in criminal robustness. Notably, the network dynamics reveal a nuanced dependence on link density, as evident in the comparison of the lower link density (red) with other densities. A critical robustness threshold around \(c^* \approx 0.2\), demarcates a regime where the stability of the criminal organization undergoes a discernible collapse. The open circles in (a, b) indicate the systems for which the $c^*$ is close to zero caused by a lack of criminal opportunity (b). In (b), the criminal opportunity decreases with role assortativity indicating specialized skills need to be available for criminal organizations to form. Criminal opportunity is defined as the number of criminal organization existing at the end (\(t = 300\)) of each Monte-Carlo run. The results presented herein pertain to a fixed parameter \(\epsilon = 10\), utilizing a network structure with \(Z = 150\) agents arranged in a ring; further details on the experimental configuration can be found in \ref{sec:orgd43c540}.}
\end{figure*}

\subsection{Spontaneous emergence of criminal organizations}
\label{sec:org68217f2}
The stability of the non-criminal and criminal strategy intricately varies depending on the initial fraction of
criminals in the system (\cref{phase_plot}). Criminal organizations can spontaneously form in the absence of
criminals in the system (\cref{phase_plot} (a)), particularly as perceived costs of criminal action reduce beyond a critical point (or perceived benefits increase), corresponding to a movement from the blue to the red area in \cref{phase_plot} (a). 
Once they emerge, the initial levels of criminals are higher, (\cref{phase_plot} (b)), which entail higher \emph{resilience} and higher \emph{robustness}. That is, perturbations in both the fraction of criminals and the cost-to-benefit ratio have little effect on the ability to form and maintain a criminal organization. Further, as the organization is formed, uncertainty in estimating cost and benefits, measured by $\epsilon$, is naturally reduced, so the system moves into an even more robust region. 

The potential for criminal organizations to spontaneously emerge carries profound implications for the efficacy of
interventions. A plausible scenario arises when there is an extreme relaxation of prolonged periods of stringent punishment for criminal activities (resulting in a perceived cost-to-benefit ratio below the critical value), which
triggers a spontaneous upsurge in criminal organizations. 
Reverting to the original perceived cost-to-benefit ration, may not suffice, requiring much stricter interventions. 
Additionally, intervening in the decision error may be a more impactful strategy than merely augmenting the cost of punishment for a crime.

\subsection{Link density and disassortatvity facilitate robustness and recruitment}
\label{sec:org92fd4a0}
The robustness of a criminal organization is contingent on its social connections and the accessibility to the roles necessary for the organization to function.
We evaluate these based on the effect of link density of the social network and the role assortativity on the dynamics. Importantly, we define criminal robustness for fixed $\epsilon = 10$ as the maximum cost-to-benefit ratio ($c^*$) such that the system dynamics sustains a highly criminal state (see \ref{sec:org627328e} for further elaboration on the methods). 
Graphs ($N=100$) are initially generated such that $Z = 150$ agents are connected in a ring with minimum role assortativity. Link density was increased by adding edges between agents having differing roles (ensuring minimum role assortativity). For each link density, the role assortativity was increased by swapping the roles of a pair of agents.

Increasing link density is associated with an increase in the maximum criminal robustness \cref{link_plot}. For a given role assortativity, $c^*$ is higher when the link density is higher. 
Increased link density provides more opportunities for criminal organizations to form since it increases the likelihood of connecting to the expertise needed to form a criminal organization \cref{link_plot}. 
Agents can connect with more individuals possessing the necessary roles, creating social opportunities for criminal organization formation or influencing citizens to join a criminal organization. 
For example, in \ref{sec:org346a4ff}, the effect of link density is studied on the likelihood of recruiting a licit community into a criminal organization.
As the link density between a criminal and non-criminal group increases, so does the likelihood of recruiting the group of non-criminals into the criminal organization (\cref{compromising_group}).

Enhancing role assortativity has a pronounced impact on diminishing the overall robustness of criminal networks. This outcome is anticipated, as a deficiency in the requisite roles prevents establishing cohesive criminal organizations. Nonetheless, the rate at which robustness decreases is noteworthy. Specifically, the increase of role assortativity within the social network restricts the formation of criminal organizations in a linear fashion until a critical point is reached, approximately at $c^* \simeq 0.25$, where a sudden shift in maximum criminal robustness is evident. This shift aligns with a seemingly minimal threshold of criminal connections required.

Illustrated in \cref{assort_plot}(b) is the computation of criminal opportunities at the conclusion of the simulation. Criminal opportunities denote the total number of potential criminal organizations an individual could form with their immediate neighbors, given their roles. The immediate drop of criminal robustness (\cref{assort_plot}(a)) can be explained by the lack of available criminal neighbors (\cref{assort_plot}(b)). Increasing the cost $c$ puts a higher requirement on the number of criminal contacts for an individual. It becomes increasingly less likely for a criminal organization to succeed with increasing cost, creating an abrupt dissolution of criminal activity for a given decision error -- in \cref{assort_plot} empirically seen as around 10 criminal neighbors.

\subsection{Increasing criminal awareness: a catalyst for the formation of robust criminal organizations}
\label{sec:orgc719ec6}
A successful criminal actor is able to thrive by forming collaborations with other agents. This requires awareness of other criminals in the system. Social networks put natural bounds on the visibility of other criminal actors. The success of forming a criminal organization is not only dependent on the immediate connectedness of an actor but also on the strategy of agents at further distances way which make their decision on their local information.

\begin{figure}[ht!]
\centering
\includegraphics[width=.9\linewidth]{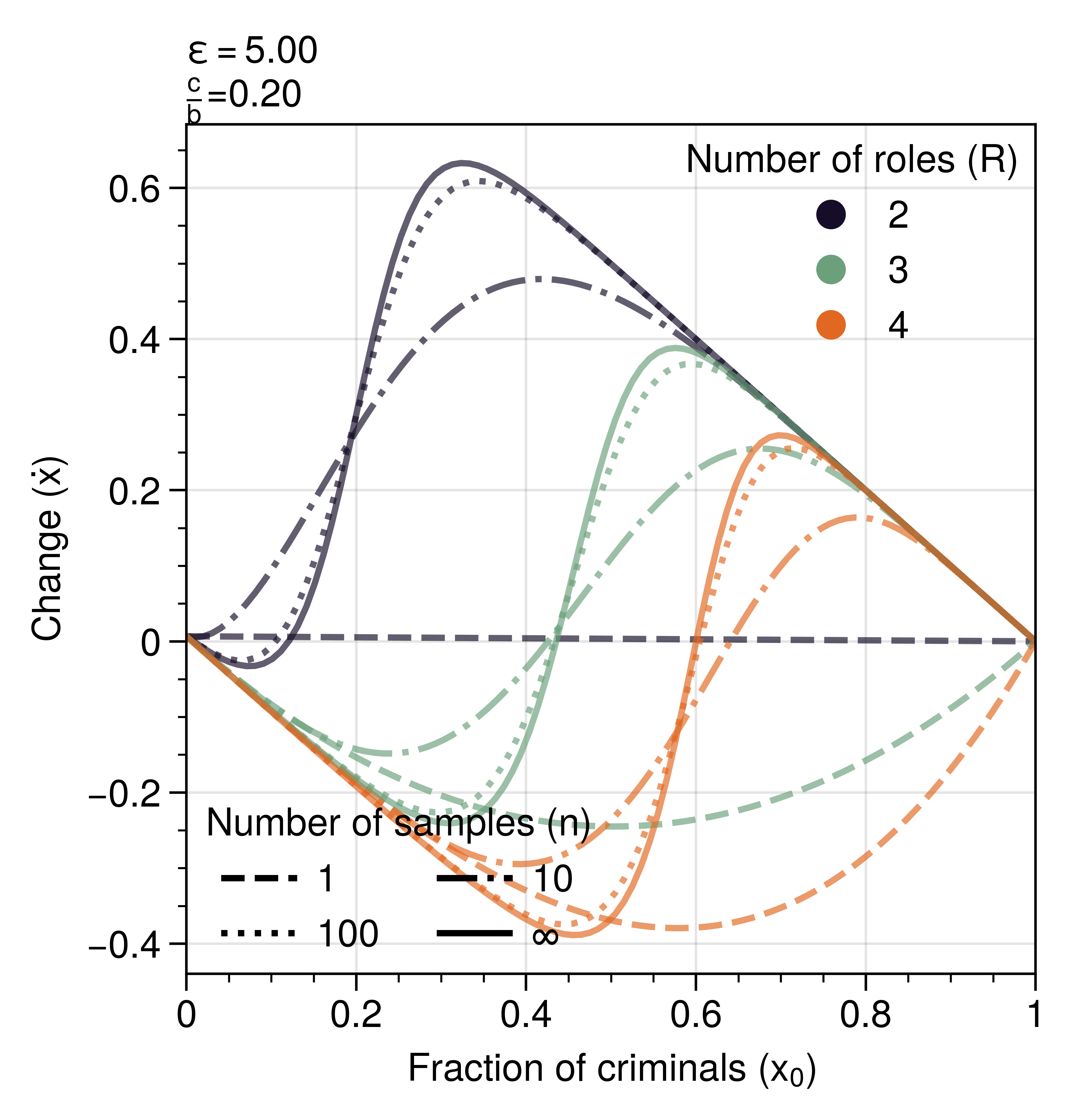}
\caption{\label{sampling_effect}The stability of the system is affected by exploration rates of the agents. Two main effects can be observed. First, for a given number of roles, criminal organizations can emerge as the agent samples its environment more. As the number of samples approaches infinity, the emergence of a criminal organization emerges; the unstable point of a forming a criminal organization is flanked by an unstable point acting as a threshold on the required initial fraction of criminal agents. Second, the result highlight that higher number of required roles imply a harder to form organization. As the number of roles increase, the unstable point shifts to the right.}
\end{figure}

The effect of access to the criminal state of surrounding forms a catalyst for forming a criminal organization. In \cref{sampling_effect}, the effect of local awareness is shown for different organization sizes. Whether a criminal organization can form is therefore a function of both the actual criminal in society, \emph{and} their visibility. When a criminal actor has low visibility, the system is characterized by a stable point at a fraction of zero criminals. If awareness increases, the criminal strategy becomes favorable if sufficient criminals are accessible to the agent. Consequently, the connectedness of a criminal will affect how likely this individual will participate and form a criminal organization; increased link density will promote the probability of criminals connecting and forming (new) criminal organizations.

\section{Discussion \& conclusions}
\label{sec:org2174d6d}
Network analysis provides an intuitive approach to decompose the structure of a criminal organization in terms of security and efficiency. However, the results of this article highlight the need to move beyond mere structural analysis and consider the intricate relationship between the organizational structure and the emergence, as well as the stability of criminal organizations amidst external conditions and interventions. Exploring the interplay between cost-benefit analysis and decision-making errors provides insights into how criminal organizations maintain resilience and robustness amidst external perturbations. These findings complement existing empirical research in criminology by providing underlying mechanisms driving the proliferation of illicit enterprises and informing strategic approaches for combating organized crime \cite{Duijn2014a,Duijn2016,Duxbury2019,Bunt2014,Morselli2007,Morselli2009a,Kleiman2009}.

Our results highlight a strong hysteresis effect in the formation and perpetuation of criminal organizations, indicating that higher resilience and robustness are achieved once a criminal organization has formed. In the criminal context, increased costs can be viewed from the standpoint of stricter punishment, either through legislative measures or intensified law enforcement efforts, or from the perspective of the financial gains a criminal organization may derive from illicit activities.

The relationship between stricter punishment and its impact on organized crime is complex and subject to debate among experts \cite{Jervis1997, Raskolnikov2020, Pratt2017, Kleiman2009}. While deterrence theory suggests that stricter punishment can reduce criminal behavior \cite{Tomlinson2016}, its effectiveness in deterring organized crime is not always clear-cut due to the sophisticated strategies employed by criminal organizations, such as bribery, corruption, and intimidation \cite{Reuter1983}. Moreover, the lack of effectiveness of increased punishment on organized crime may be attributed to the creation of in-group versus out-group dynamics that occur within criminal organizations. In such circumstances criminal co-offence reduces the impact of external perturbations and increases their resilience and robustness \cite{vonLampe2004, VonLampe2016}.

Increased punishment is not the sole determinant of the effectiveness of a criminal organization \cite{Morselli2011}. Depending on the diversity of the portfolio of the criminal organization, the benefits derived from engaging in criminal activities are contingent upon the forces of supply and demand \cite{Schelling2023}. These forces may compel criminal organizations to alter their criminal portfolio, either by moving into different financial markets or by changing their methods of operation \cite{Albanese2020, Ehrlich1996}.

The insights gained from this study have important implications for policymakers and law enforcement agencies. They can be used to inform the development of targeted intervention strategies that address network resilience, awareness dynamics, and societal factors driving illicit behavior. The results reveal how interventions of similar scale may lead to the downfall of a criminal organization under certain conditions, while delivering marginal effects under others. This implies that increasing criminal punishment may not have the desired effects on reducing organized crime, while having a profound effect on the personal freedoms novel policies may entail \cite{Datta2017, Hagan1999, Calland2000}. Understanding the cost-benefit analysis and decision-making processes within criminal organizations is imperative for formulating effective policy frameworks aimed at preventing and disrupting organized crime \cite{Manning2010, Tyler2002}.

Furthermore, this perspective reframes law enforcement policies not merely as reactive measures to criminal actions but as integral components of a dynamic system. Within this system, policymakers, law enforcement agencies, and criminal organizations form a collective wherein each party can influence the other. This systemic approach underscores the interplay between enforcement strategies and criminal behavior, emphasizing the need for proactive, adaptive policy measures that anticipate and counteract criminal activities effectively \cite{VanCalster2006, vanCalster2015, Duijn2014a, Duijn2016}.

Looking ahead, there remain several unanswered questions and areas for future research. Further exploration of temporal dynamics, environmental adaptability, and the nuanced dynamics of social ties within criminal networks is warranted. The literature underscores the social embeddedness of criminal organizations, and introducing heterogeneous agents to include a chain-of-command could enhance the realism of the criminal process. Additionally, specifying different sources of cost or benefit could provide insights into how the effects of law enforcement and the supply and demand of illicit goods interact and affect the effectiveness of criminal organizations. Moreover, integrating trust between agents or dynamics between roles may lead to a more detailed understanding of real-world criminal organizations.

Finally, the integration of real-world data to validate and extend model findings, coupled with an examination of intervention strategies targeting structural dependencies, will contribute to a more comprehensive understanding of effective crime prevention measures. In sum, this study offers valuable insights into the dynamics of criminal organizations, laying the groundwork for future research in this field.

\section{Conflict of interest}
\label{sec:orgc51fa29}
The authors declare no conflict of interest.
\section{Author contribution}
\label{sec:org3a86c5c}
\textbf{Casper van Elteren}: initial draft, analysis, idea formation.
\textbf{Vítor V. Vasconcelos}: review draft, analysis, idea formation.
\textbf{Mike Lees}: review draft, analysis, idea formation. 
\section{Funding}
\label{sec:org04a7497}
This research is supported by grant Hyperion 2454972 of the Dutch National Police.
\printbibliography

\section{Code and Data availability}
The code is publicly available at https://github.com/cvanelteren/boiler_room. The data can be re-generated through te provided scripts.
\newpage
\appendix
\counterwithin{figure}{section}

\section{Numerical Methods}
\label{sec:orgd43c540}
\subsection{Graph generation and link density, and assortativity}
\label{sec:orgd802cfb}
For the Monte-Carlo results, graphs were generated by
spawning a ring with \(Z=150\) in which each role was
alternating ensuring that the base graph would have maximum
assortativity, with minimum degree. Furthermore, this
ensured that each role occurred equally (\(Z_r = 50 \forall r \in R =
1, \dots 3\))

The link density was increased by adding new edges between
agents of non-matching roles. That is, only edges were added
if the two roles did not share the same role. This would
ensure that the assortativity for the base graph would not
increase with the number of edges added. the link density
was increased from the base ring by a fraction \(f \in \{0,
0.1\}\) in 5 equally sized steps. This range was determined
by first evaluating up to \(f=1\), and seeing that above 0.1,
the values don't increase much further.

\begin{figure}[htbp]
\centering
\includegraphics[width=.9\linewidth]{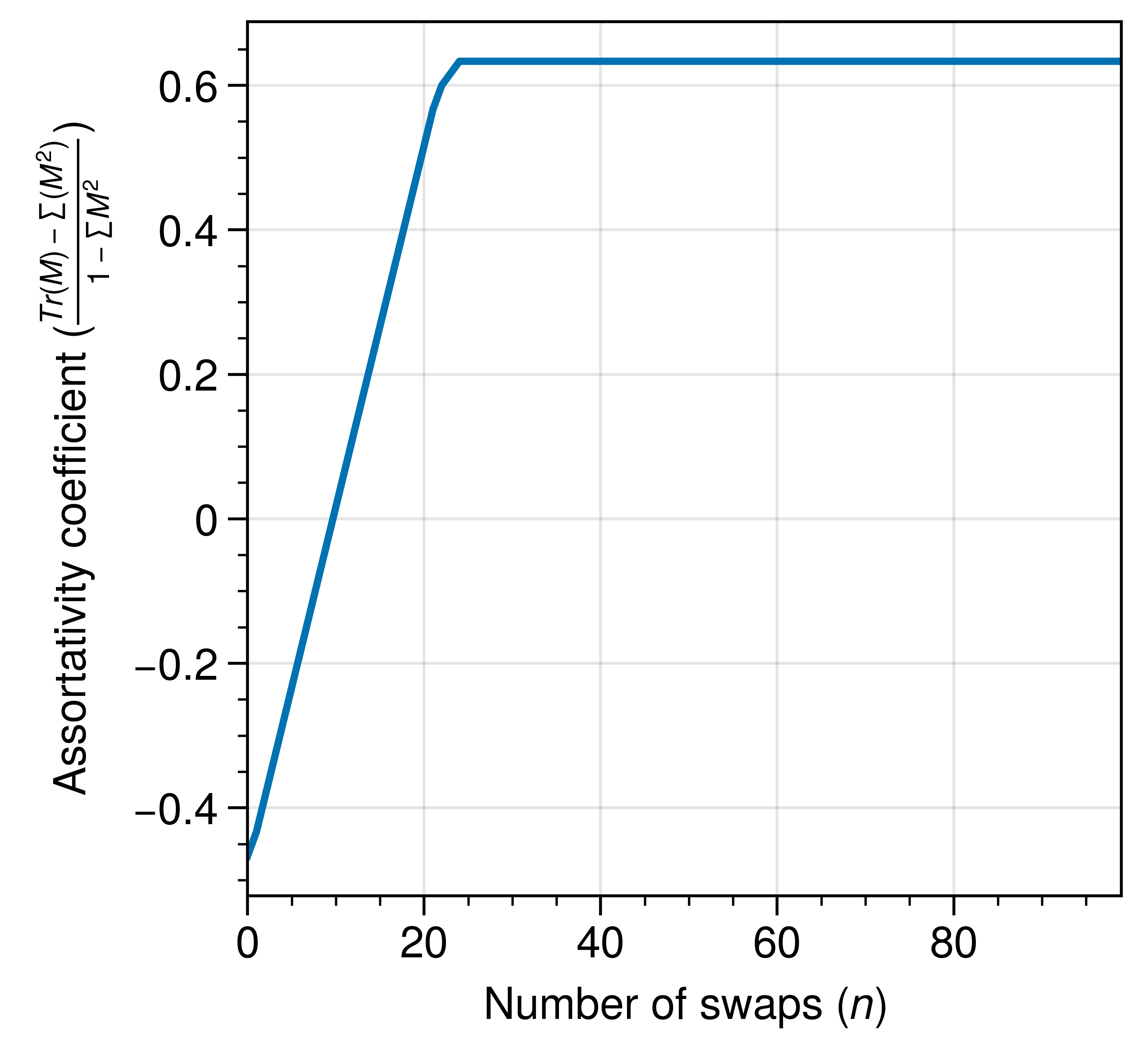}
\caption{\label{assortativty}A typical output for increasing assortativity for fixed link density; as the number swaps increase, the associativity increases.}
\end{figure}

Assortativity was increased by considering \(n\) swaps such
that the assortativity \(q\) would increase to \(q'\), this was
performed using exhaustive search. Only graph edits were
allowed that increased the assortativity. The number of
swaps was varied between the range \([0, 100]\) with dynamic
step size. The range \([0, 10]\) was sampled densely, after
the number of swaps increased by increments of 5 (until 20),
and 10 (>20). The assortativity coefficient for graph \(G\)
for the roles as attributes as:

\begin{equation}
\label{assort}
A_G = \frac{Tr(M) - \sum(M^2)}{1 - \sum M^2}
\end{equation}
where \(M\) is the joint probability distribution or mixing
matrix \cite{Newman2003}.
\subsection{Monte-Carlo simulations}
\label{sec:org61463f1}
The effect of link density (\cref{assort_plot}) were performed
using Monte-Carlo methods. The maximum incurred costs
(\ref{sec:org627328e}) was computed using bisection search.
This was achieved by starting the simulation from a state
were all agents were criminal or non-criminal and then
stepping the system for \(T = 1000\) steps. The final state
was used to estimate the fixed point by computing the
average state of the nodes in the system. Each simulation
step, consists of updating all the nodes in the system (\(Z =
300\)) in random order.

The model was written in Nim (version 2.0). Random numbers
were generated using the xoroshiro128+ generator provided
by Nim's standard library. Visualization and data-analysis
was performed in python using Matplotlib. Source code is
available at \url{https://github.com/cvanelteren/boiler\_room}.

\subsubsection{Numerical validation of phase diagram}
\label{sec:org88d1d9c}
\begin{figure}[htbp]
\centering
\includegraphics[width=.9\linewidth]{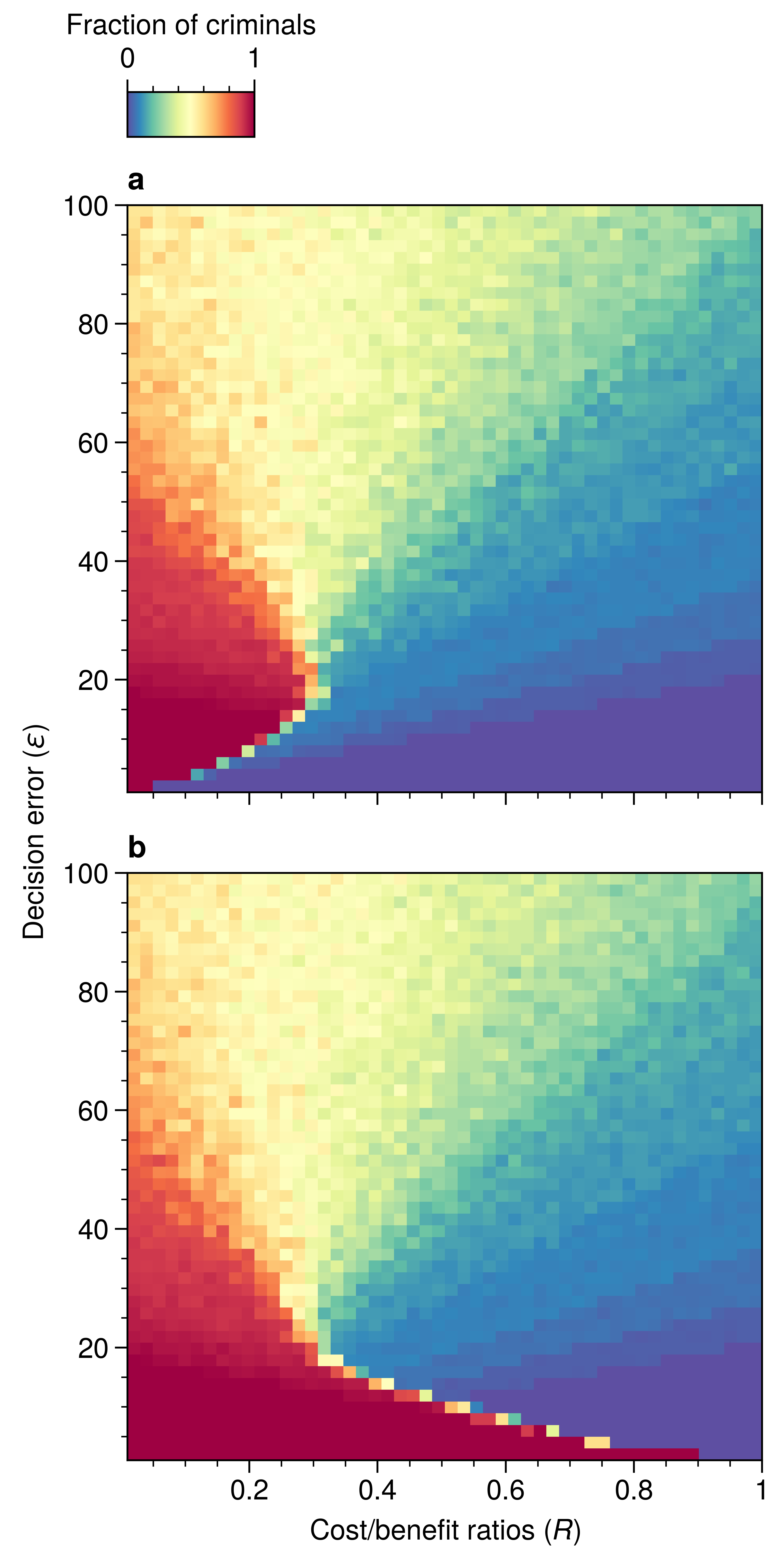}
\caption{\label{mc_well_mixed}Simulations were performed on a complete graph with \(Z=150\) agents. For \(t=300\) time steps the simulations were run and repeated for \(T=100\) trials. Shown are the system averages as a function of decision error (\(\epsilon\)) and cost/benefit ratio \((R)\).}
\end{figure}

Using the methods described in
\ref{sec:orgd43c540}, we replicated the
results analytical results shown in \cref{phase_plot}. For the
Monte-Carlo simulations we choose the number of local samples to
be \(n = 100\) for each agent which is shown to converge
closely to the analytical solution (\cref{sampling_effect}).
The Monte-Carlo results on complete graphs show close
similarity to the analytical result(\cref{mc_well_mixed})

\subsection{Criminal robustness: incurred cost}
\label{sec:org627328e}
We defined maximum incurred cost as the maximum cost for
which a criminal organization can be formed. For a given
noise value \(\beta\), we hypothesize that the networks can
nurture or deprive conditions to form criminal
organizations. Let \(S^t = \{s_1, s_2, \dots, s_z\}\) indicate
the (criminal) state of the agents at time \(t\), then we can
write the criminal \emph{robustness} by its maximum incurred cost
as

\begin{equation}
\label{maximum_incurred_cost}
c^*(\epsilon, \langle S^0 \rangle = 1) = \max_c \langle S^t\rangle \neq 1.
\end{equation}

We are interested at which cost \(c^*\) the system has only 1
stable attractor. In other words, at which cost, does the
system loose its ability to sustain a saturated criminal
market.

The maximum incurred cost is evaluated for networked systems
numerically through bisection search. Simulation were
performed starting with all agents in the criminal state \(\langle
S^0\rangle = 1\). Since the maximum inccured cost depends on the
noise level, care was taken into choosing a proper value
such that there exists a stable point for a saturated
criminal market. The result from \cref{phase_plot} was used to
determine a value for \(\beta\). The analytical results form an
upper bound for networked systems. That is, the network
structure will impose constraints on the ability to reach a
saturated market for a given noise level; it cannot improve
over the well-mixed conditions As such it was determined
that \(\epsilon = 10\) was used to compute \(c^*\) for the networked
systems.

The simulation time was set to \(t=300\) using the same update
settings as described in
\ref{sec:org61463f1}.
\subsection{Criminal opportunity}
\label{sec:orgbf00c19}
Criminal opportunity is defined as the instantaneous number
of criminal organization out of all possible organizations
at the end of a simulation, or

\begin{equation}
\label{criminal_opportunity}
O^t = \sum_{i=0}^Z \frac{M_i^t}{\mathcal{M}_i^t},
\end{equation}

where \(M_i^t\) are the number of criminal organizations for
agent \(i\) at time \(t\) and \(\mathcal{M}_i^t\) indicates the total
organizations agent \(i\) could have formed at time \(t\).

\section{Derivation of \cref{payout}}
\label{sec:org28cfbc4}
\Cref{payout} describes the payout of a singular game. We
consider the change of the three subpopulations as a
stochastic variable. Let \(x_r\) describe the fraction of
criminal of each role \(r \in R\) in the population, then the
expected payout \(r\) becomes

\begin{dmath}
\label{deriv:population_payoff}
 \pi_r = (1 - x_r) \sum_{k = 0}^n (\prod^R_{j=0, j \neq i} x_{j})^k (1 - \prod^R_{j=0, j \neq i} x_{j})^{n - k}
 \frac{1}{1 + e^{-\frac{1}{\epsilon}(kb \prod^R_{j=0, j \neq i} x_{j} - cn)}}.
\end{dmath}

Following, we can describe the instantaneous change of the
system as the balance between the gains and losses of a
criminal in a particular role \(x_r\). We can write:

\begin{dmath}
\label{full_change}
\footnotesize
\frac{d}{dt} x_i = \mathbb{E}[x_i^+] - \mathbb{E}[x_i^-] = (1 - x_i)
\sum_{k=0}^n \binom{n}{k} \bigl(\prod_{j=0, j \neq i}^R x_j\bigr)^k \left(1 - \prod_{j=0, j \neq i}^R x_j\right)^{n-k} \frac{1}{1 + e^{-\epsilon (\frac{k}{n} b - c)}}
- x_i \sum_{k = 0}^n \binom{n}{k}\left(\prod_{j=0, j \neq i}^R x_j\right)^k \left(1 - \prod_{j=0, j \neq i}^R x_j\right)^{n-k} \frac{1}{1+e^{-\epsilon(\frac{k}{n} b - c)}}
= \sum_{k = 0}^n \binom{n}{k} \left(\prod_{j=0, j \neq i}^R x_j\right)^k \left(1 - \prod x_{j \neq i}\right)^{n - k}\big(\frac{1 - x_i}{1 + e^{-\epsilon (\frac{k}{n} b - c)}}
- \frac{x_i}{1+e^{-\epsilon (\frac{k}{n} b - c)}} \big)
= \sum_{k=0}^n \binom{n}{k}\left(\prod_{j=0, j \neq i}^R x_j \right)^k \left(1 - \prod_{j=0, j \neq i}^R x_j \right)^{n - k}
\frac{1}{1+e^{-\epsilon b (k - \frac{c}{b}n)}} - x_i.
\end{dmath}

In the limit of \(n \to \infty\), section \cref{full_change} can be written
as
\begin{equation}
\begin{split}
 \lim_{n\to\infty} \frac{d}{dt} x_i = \frac{1}{1 + e^{- \frac{1}{\epsilon} b(\prod_{j = 0, j \neq i}^R x_j - \frac{c}{b})}} - x_i,
\end{split}
\end{equation}
see \cite{Bernstein1912} for proof.

The model formulation is akin to a n-order Ising model
with base \(\{0, 1\}\) under transversal magnetic field where
each agent takes local averages \cite{Simon2009}.
\section{Role diversification drives bifurcation}
\label{sec:orgc0093f7}
\begin{figure}[htbp]
\centering
\includegraphics[width=.9\linewidth]{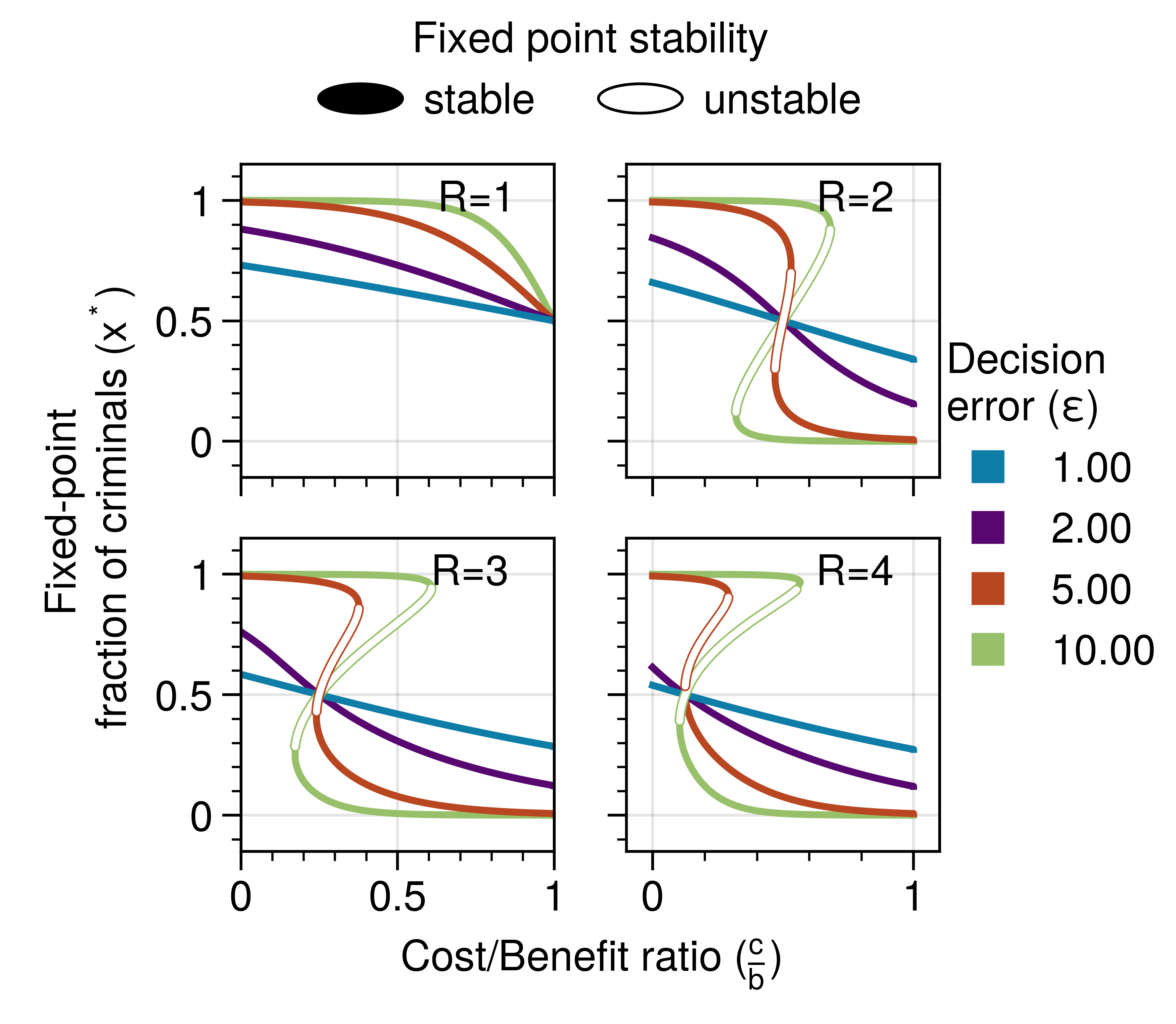}
\caption{\label{bifurcation_per_n}Bifurcation diagram of the system dynamics as a function of the number of roles. From left two right the organization consists of 1, 2, 3, and 4 roles respectively. Without different roles (n=1), there exists a net benefit to become criminal as the payout outweighs the cost regardless of the size of the cost. As the number of roles grows, two stable equilibria emerge. There exists, a range of values of the cost/benefit ratio for which the a complete criminal organization, and non-criminal strategy are valid. No in between strategy exists. As the number of roles grows, the region of cost/benefit ratio where there exist three stable points grows}
\end{figure}

For lone wolf criminals (\cref{bifurcation_per_n}, top left),
\eqref{limiting_change} implies the existence of a singular
stable point as a function of the benefit ratio and given
noise value. Noise affects the dynamics of system by
sharpening (or soften) the sharpness of the curve. For high
noise conditions, criminal strategy becomes random (50/50)
whereas for lower noise values, the criminal strategy
becomes favorable for high benefit \(b\).

For \(R \ge 2\), a bifurcation emerges. As the cost decreases
past a critical point (\(\frac{c}{b} \approx 0.5\)), an unstable
fixed point occurs flanked by two stable fixed point; one in
which a high level of criminal activity is present, and
another in which the criminal activity vanishes from the
system. The noise parameter (\(\epsilon\)) controls the location
of the unstable point. As \(\epsilon \to 0\), the fraction of
criminals required to evolve towards a complete criminal
system decreases. That is, fewer criminals in the system are
necessary for the system to evolve towards the stable
attractor where the majority of individuals are criminal.

Increasing the number of roles above 2 has two major
effects.

First, it becomes more difficult to form a criminal
organization. Consider for example \(\epsilon = 2\). For a
criminal organization consisting of 2 roles, there exists
only 1 stable attractor as a function of the cost/benefit
ratio. As the roles increase, the inflection towards
criminal activity becomes sharper, requiring lower cost (to
benefit), eventually bifurcating at \(R = 4\). In general, the
minimum required cost is shifted to the left compared to \(R
= i + 1\) than \(R = i\) indicating it is more difficult to
coordinate forming a criminal organization.

Second, if a criminal organization forms it is generally
more robust to perturbations. The drive towards a complete
criminal system is higher for a given cost/benefit ratio if
the criminal strategy is viable. Consider \(x^*\) for \(R=0.2\),
the stable equilibrium around 1 is reached earlier as the
number of roles (\(R\)) increases. Combined with the previous
point, this indicates that if a criminal organization forms,
it it is more robust and potent than a similar organization
forming with a lower level of diversity of roles.

\section{Link density facilitates recruitment into a criminal organization}
\label{sec:org346a4ff}
\begin{figure}[htbp]
\centering
\includegraphics[width=.9\linewidth]{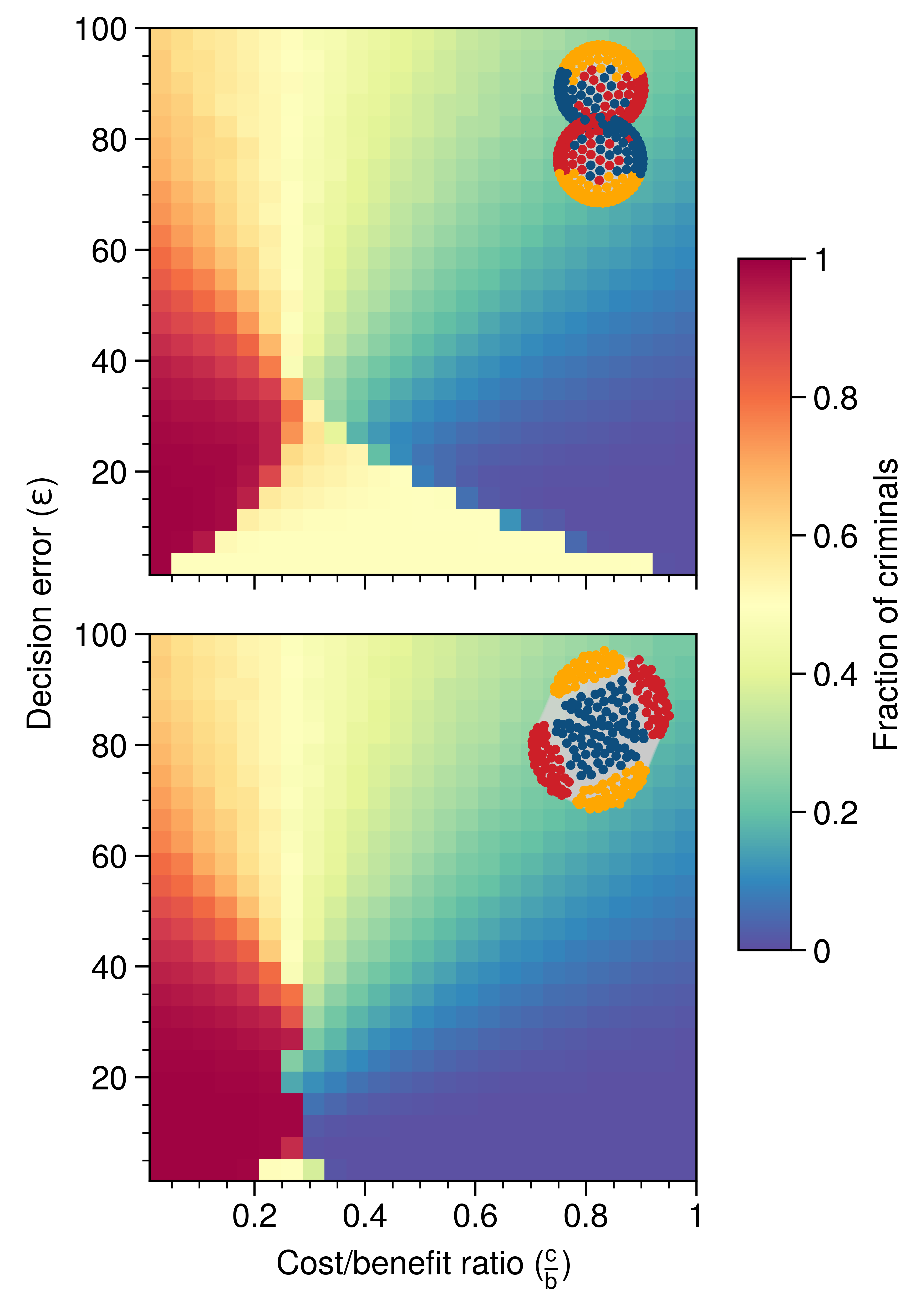}
\caption{\label{compromising_group}Non-criminal groups can be comprimised by increased link density between the groups. Shown are Monte-Carlo results from two groups that share no social interactions (a) and those that are tightly knit. The inset illustrates the two communities where the node color indicates different roles essential to the organization. (a) When the criminal group is separated from the non-criminal group, the criminal \emph{robustness} shows a heavy tail in which only the criminal group remains criminal. (b) When the link density between the groups increases, the criminal \emph{robustness} (red area) expands further than when the communities were separated. This indicates that the criminal organization was able to (i) compromise the non-criminal group, and (ii) create a more robust criminal organization.}
\end{figure}

We investigate the importance of these indirect connections
by generating two organizations, one criminal and one
non-criminal. By varying the connectivity between the two
organizations, we expect that there exists some threshold
value of cost/benefit and noise for which the non-criminal
organization flips to criminal activity. In
\cref{compromising_group}, we observe that the
non-criminal organization spontaneously converts to a
criminal organization for sufficiently high noise and high
benefit. This effect occurs earlier when the number of
connections with the criminal organizations increases
(\cref{compromising_group}). The increased
connectivity between the two organizations shows how a
greater degree of ties between under and overworld can
effect the propagation of criminal behavior.

\section{Theories of organized crime}
\label{sec:org2890a8a}

\begin{quote}
\emph{I shall not today attempt further to define the kinds of material I understand to be embraced within that shorthand description. But I know it when I see it, and the motion picture involved in this case is not that
I know it when I see it} - Justice Potter Stewart \cite{zotero-26142}
\end{quote}

Understanding crime can be tackled from multiple
perspectives. A theory on crime provides a lens to view the
topic. Crime, like any abstract topic, is difficult to
define yet easy to recognize. The concept of crime touches
upon many different topic such as ethics, morality, the rule
of law, philosophy, and psychology to name a few. To merely
list the plentiful theories of crime would make the
discussion that follows too broad and unbearable. I
therefore want to focus on a narrower topic: organized
crime.

Organized crime differs itself from other forms of crime by
its adjective. The crime is not merely any crime, \emph{it is
organized}. The addition of this word invokes imagery of
motor gangs, Mafia, and cartels, but less so of the
ordinary thief. However, organized crime usually refers to
something more than merely a criminal act that was
performed. A burglary that is prepared is therefore not
necessarily considered to be organized crime. What then is
this \emph{organized crime}?

Why is it important to define organized crime? Organized
crime encompass many different aspects of society that are
difficult to grasp in its entirety. Surely any definition of
crime will fall short of what organized crime is
\cite{Hagan2006,Schelling2023}. The concept of what is
considered criminal may change over time demanded by changes
in society. However, a definition will form a stepping stone
towards understanding the problem that law enforcement aim
to reduce. That is, by discussing the topic, and reviewing
different perspectives, we aim to come closer to a truth at
least temporarily. The truth is not in the sense of some
Platonian Idea, but rather a workable form that describes
adequately the topic of interest.

A definition may aid in the formation of new policy and or
laws, and is therefore crucial for computational scientists
to align ourselves with a form that warrants computational
investigations. It prevents ``conceptual puddings'' where
stereotypes gain a strong footholds and forces us, the
scientists, to center around assumptions that inform our
models.

After delving into the etymology I will discuss theories
influencing the current work. See for a more comprehensive
overview \cite{Bunt2014}.

\subsection{Etymology of organized crime}
\label{sec:org5d5607d}
The term ``organized crime'' began to gain prominence in the
early 20th century, particularly in the United States. It
became more commonly used during the Prohibition era, which
lasted from 1920 to 1933, when the production, sale, and
distribution of alcoholic beverages were prohibited by the
18th Amendment to the U.S. Constitution.

During this period, criminal organizations, such as the
Italian-American Mafia and other groups, were heavily
involved in illegal alcohol production and distribution, as
well as other criminal activities. The activities of these
criminal organizations and their structured, coordinated
nature led to the popularization of the term ``organized
crime'' to describe their activities.

Since then, the term has been used to refer to various forms
of criminal enterprises and activities that involve
structured, coordinated, and often transnational criminal
organizations. It has become a widely recognized concept in
criminology, law enforcement, and legal discussions.
\subsection{Illicit enterprise theory}
\label{sec:orgd99caa0}
Criminal organizations show remarkable similarity with legal
organizations \cite{Bunt2014}. To commit a criminal act
often requires different skills and expertise working
together to obtain profit. Indeed some law enforcement
strategies focus on mapping the criminal value chains,
disrupting the ability of the criminal organization to
execute their activities \cite{Duijn2016,Duijn2014a}.

The value chain includes all roles or activities that
support the primary activities bringing a raw resource or
service to market \cite{Porter2011}, human resource
management, research and development, firm infrastructure
and so on. The value chain describes the business process or
coarse steps that make up the business.

For a criminal organization, may include roles or services
that do not occur in licit business. For example,
strong-arms or murders for hire could be employed to ensure
the market position of a criminal organization within the
criminal infrastructure. Similarly, logistic roles such
acquiring a getaway care may require carjackers to acquire
the vehicles.

In short, illegal enterprise theory states that criminal
organizations share more similarities than differences with
licit businesses. Members of criminal behaviors use
rationale and planning to further business endeavors.
\subsection{Social embededness of criminal networks}
\label{sec:orgee8c52d}
A network in general is used to represent data. As such it
requires a filter on the meaning of the nodes and edges in
the network. For a social network this could be contact,
kinship, geographical distance and so on. The interpretation
and meaning of social network analysis is therefore highly
context dependent.

For a given representation of a social network, other
subnetworks can be identified. Consider, for example, a
network formed from contact data. Within this social network
different subnetworks can be identified. The network of work
colleagues is \emph{embedded} on this social network, similar to
friend networks, or kinship networks. The concept of social
embedding gains its origins from social economy
\cite{Granovetter1985}. Sociologists identified that economic
transactions are embedded within the networks of personal
relations. Criminal networks, therefore, are another form of
embedding \cite{Bunt2014}.

The general idea is that criminal opportunity is formed
through social relations. Social relations do not happen at
random but often abey the laws of social an geographic
distance \cite{Feld1981}. The closer people live, the more
daily activity people have in common, the less social
distance between them. This results in a clustering of
people based on factors such as geographic distance,
ethnicity, education, age, membership of a local sports club
and so on. This aspect is also found in criminal networks
\cite{Scott2011a}.

The important point is that for a given observed network
structure, the structural position influences the
embededness of an individual to be exposed to criminal
activity. The network is not a final state, but the snapshot
of the outcome of a complex adaptive process that
continuously evolves. To understand how some networks thrive
while others falter, requires understanding the relationship
between the individuals and the kind of dynamics that exists
between them.
\end{document}